# Light Absorption in Arctic Sea Ice – Chlorophyll and Black Carbon

O. Ogunro[1, 3, *], S. Elliott[2], N. Jeffery[2], F. Hoffman[3], H. Wang[4], O. W. Wingenter[1]

[1] Department of Chemistry, New Mexico Tech, Socorro, NM 87801, USA
[2] Los Alamos National Laboratory, Los Alamos, NM 87545, USA
[3] Oak Ridge National Laboratory, Oak Ridge, TN 37831, USA
[4] Pacific Northwest National Laboratory, Richland, WA 99354

[*] Corresponding Author: O. Ogunro ogunrooo@ornl.gov

Key Points
- Arctic sea ice loss
- Light absorption by black carbon deposited on Arctic snow/ice
- Light absorption by chlorophyll within and beneath Arctic ice pack

**Abstract**

Arctic sea ice extent has declined continuously for the past decade, owing partially to light absorption by black carbon (BC) and other impurities deposited on snow and the underlying pack. We present simulations for the contemporary period showing that the optical depth contributed by Arctic ice algal chlorophyll may be comparable during Boreal Spring to the corresponding values attributable to BC. The largest chlorophyll attenuation is obtained in the bottom layer, which supports pigment concentrations of about 300 – 1000 mg/m$^3$ in the Bering Sea and Sea of Okhotsk. However, chlorophyll concentrations for the ice interior in regions north of 75° N and across the Canadian Archipelago are less than 0.1 µg/m$^3$. Freeboard and infiltration communities lead to intermediate levels of light removal. Since BC works its way downward from the atmospheric interface, there will be regions where it appropriates photosynthetic capability. Where ice thicknesses permit significant penetration through the pack column the ice algae may be crucial absorbers. We propose a continuous increase in relative chlorophyll activity and attenuation in the future, as biological activity becomes stronger in thin ice toward the center of the Arctic basin. A shift in relative importance of the two absorber types could occur as total BC mixing ratios are reduced because of environmental advocacy.

**Introduction**

Arctic sea ice represents an essential component of the global climate system because of its influences on planetary albedo, ocean circulation and high latitude atmospheric insolation-insulation [Aagaard and Carmack, 1989; Piexoto and Oort, 1992; Mauritzen and Hakkinen, 1997]. The ice pack helps to maintain the stability of vast marine methane deposits, and it also regulates biology activity in support of complex Arctic food webs [Shakhova et al. 2010: Arrigo et al. 2010]. Significant photosynthetic activity ongoing within/below the ice column influences natural production and emission of climate active gases such as dimethylsulfide and $CO_2$ [Stefels et al. 2007; Trevena and Jones, 2012; Tison et al. 2008; Rysgaard et al. 2011]. The potential for polar amplification of climate change points strongly to a need for improved understanding of the Arctic biogeochemical system and its epontic component. A major concern will be the influence of absorbing agents on radiation transfer, whether the substances in question are deposited onto or generated within the pack.

The bright surface of snow and ice at high latitudes constitutes a medium of high albedo that can reflect about ninety percent of incident shortwave radiation. In contrast to this high surface reflectivity at the poles, solar radiation absorption is higher in mid-latitude, primarily because surfaces are often covered in vegetation or open ocean [Herman et al., 2001; Herman et al., 2009]. The high albedo of bright-white snow and ice at the north pole makes the Arctic environment more susceptible to the optical influence of impurities. Presence of these impurities at snow/ice surface may thus impact surface reflectivity by increasing the rate of frozen water loss during the melting season. Ultimately this effect can amplify global warming [Warren 1982; Doherty et al., 2010].

The fingerprint of climate change is more obvious in the Arctic than any other place on Earth. This is not only because the surface temperature there has increased at twice the rate of the global mean, but also because Arctic sea ice extent has reached a record low 49% reduction relative to the 1979-2000 climatology [Overland et al. 2013]. The Arctic is extremely sensitive to change because of the potential for a large drop in surface albedo. Vast and highly reflective regions of Arctic sea ice become more absorptive as they convert to open water [Barnhart et al. 2015]. This transition forms a positive feedback loop. Among other effects, it has therefore increased the temperature of on-ice air flow within the polar boundary layer [Brümmer and Thiemann 2002; IPCC 2007].

Models representing current state of global climate have shown considerable difficulty simulating variations in the rate of Arctic sea ice disappearance [Stroeve et al. 2007; Serreze and Barry, 2011; Stroeve et al. 2012; Pithan and Mauritsen, 2014]. Observations and in situ measurements of ice melt and thickness diverge from computations, with models usually under-predicting sea ice loss. Datasets collected from different sources (ship, aircraft and satellite) indicate that September ice coverage has fallen at a rate of 7.8 percent per decade during 1953-2006 period as against an average of 5.4 percent per decade reported by the ensemble of models [e.g Stroeve et al 2007]. This implies that conservative predictions offered by current generation simulations need to be improved and optimized. It is possible that there are unaccounted forcings or feedbacks which contribute significantly to the intense reductions. Increased vegetation within the Arctic Circle can be attributed to habitat migration, and it plays a significant role in regional scale climate evolution. But the reduction in sea ice extent may have a similar outcome, opening the way for increased biological activity both alongside and within Arctic sea ice. Iron is readily available in Northern waters [Lannuzel et al. 2010; Wang et al. 2013b] along with the major macronutrients [Hydrological Atlas of the Arctic Ocean]. Early blooms are likely in peripheral polar seas and high latitude coastal areas. In fact the combined open water and ice algal communities support the highest chlorophyll concentration of any marine environment [Arrigo et al 2010].

Various mechanisms have been put forward to explain the climate transformations ongoing in the contemporary Arctic. These include alterations in cloud cover, an increase in atmospheric water vapor, atmospheric and oceanic heat transport from lower latitudes and also the various forms of strong surface albedo feedback [Holland and Bitz 2003; Alexeev and Jackson 2013 and more]. Careful consideration of these lines of reasoning suggests that decreased sea ice extent is a major driver for Arctic amplification, operating through the positive albedo feedback [Screen and Simmonds, 2010]. Although models are now predicting nearly ice free summer conditions for the Arctic Ocean beginning in about 30 years [Stroeve et al. 2007; Overland and Wang, 2013], it is clear from observations that Arctic coverage is near the minimum value ever recorded and it is still decreasing continuously. Simulations of the processes contributing to melting dynamics must improve in order to reasonably represent this critical system.

Considering the strong potential feedback of Arctic pack ice on global climate and upcoming environmental change, it is crucial to investigate dynamic factors that could influence the polar albedo feedback. Radiation absorption by the industrial black carbon (BC) deposited on Arctic snow and sea ice surfaces is one hypothesized contributor to the coverage decline [Flanner et al. 2007; Bond et al. 2013; Kerr, 2013]. However, high concentrations of the common pigment chlorophyll have been reported near and within Arctic sea ice in the Spring-Summer, due to favorable polar conditions for biological activity [Arrigo et al. 2012]. In the present work we propose to investigate optical effects of both black carbon/soot and the algal biopigments within the Arctic regime. It is shown that light absorption by chlorophyll-a generated within the ice pack may compete effectively during boreal spring.

Simulations of aerosol transport and polar biogeochemistry are now improving rapidly in Earth System Models [Elliott et al. 2012; Burrows et al. 2014; Wang et al. 2013]. Inputs of radiatively active material from above and below the regional pack simultaneously influence the sea ice energy budget, with the potential to feed back onto the regional albedo through loss of ice extent. Recently documented polar climate change amplifications may result at least in part from the combination of black carbon interaction with the cryospheric reflection plus ecodynamic effects of ice domain chlorophyll [Zeebe et al. 1996; Flanner et al. 2007; Ackley et al. 2008; Lengainge et al. 2009; Marks and King, 2013].

Our emphasis here is on estimation of absorption by the pigments, and especially chlorophyll-a, as produced by biological activity within the Arctic sea ice domain. To provide an anthropogenic perspective, we compare with the effects of black carbon deposition on the Arctic snow/ice column. In some locations a tradeoff may exist between light absorption by soot deposited on the pack and chlorophyll produced within and beneath the ice by ongoing biological activity. In the era of anthropogenic induced climate change and massive high latitude ice losses, this competition takes on special significance. The issue is particularly challenging because of high uncertainties inherent to large scale climate, atmospheric tracer and biogeochemistry simulations, and the difficulties faced by field scientists in sampling polar regions. Accurate estimates of long range aerosol transport to remote areas of the Northern Hemisphere are challenging [Wang et al. 2013a] and models of boreal marine ecosystems are yet to be fully developed [Arrigo et al. 1997; Deal et al. 2011; Elliott et al. 2012; Tedesco and Vichi, 2014].

Modeling of sea-ice physical and biological processes together with experiments and field observations promise rapid progress in the quality of Arctic ice predictions. Here we develop a dynamic ice system module to investigate discrete absorption of both BC and chlorophyll on/in the Arctic pack, using BC deposition fields from version 5 of Community Atmosphere Model (CAM5) and vertically distributed layers of chlorophyll concentrations from Sea Ice Model (CICE).

We hypothesize that in addition to black carbon deposited on Arctic sea ice and its overlying snow cover, ice algal chlorophyll is likely to act as significant absorber and redistributor of energy in at least some areas. Absorption by anthropogenic materials such as black carbon can thus be compared with that of natural pigments. We chart areas of anthropogenic versus biological absorption dominance in the sense of simple absorption and single scattering. Later we will couple this knowledge into a full radiation transfer code in order to apportion the various mechanisms to polar climate change amplification. The work prepares us to study more traditional issues such as chlorophyll warming of the pack periphery (Lengainge et al. 2009) and chemical effects of the flow of organics from ice internal communities (Krembs et al. 2011). Modeling of these regional situations can both complement experimental work and help guide future observations.

## Black Carbon and Chlorophyll-a

Black carbon is a highly (almost purely) carbonaceous material that can be generated naturally or anthropogenically through combustion in processes. These include diesel engine combustion and open biomass burning [Bond et al. 2013]. Particles are first transported vertically from the point source, then horizontally through the atmosphere. In the Northern hemisphere, a mass of black carbon released into the atmosphere could potentially be transported by wind dynamics until it makes its way to pristine ecosystems of the Arctic. Although this substance could be removed from the atmosphere through deposition or contact with surfaces, considerable amounts still move to the Arctic Circle and beyond, so that they are then deposited on the polar snow and ice [Doherty et al., 2010; Wang et al., 2013]. Since black carbon is highly absorptive in the visible spectrum, the presence of low concentration (as little as 1 part per million by weight) could lower the albedo of Arctic snow in the shortwave by about 5-15% [Warren and Wiscombe 1980].

In contrast to impurities that are deposited from the atmosphere, high concentrations of biopigments including chlorophyll-a are produced inside the ice pack by sea ice algal assemblages. The pack ecosystems are distributed non uniformly in vertical layers [Ackley and Sullivan 1994; Arrigo et al. 2010]. Environmental and thermodynamic conditions applying to each slab of biological activity determine the amounts of light, space and heat energy available plus the nutrients required for primary production [Arrigo et al 2003]. Convection in the brine channels of columnar ice supports upward flow of nutrients, which leads to primary production in the presence of even moderate light levels [Thomas and Papadimitriou 2003]. The chlorophyll-a produced during this process is, roughly speaking, just as absorptive in the visible as soot (compare coefficients). Ongoing thinning of sea ice and subsequent opening of melt water ponds in the Arctic has created an avenue for enormous increases in the strength of algal blooms. Ultimately the associated pigments must increase the total sea ice absorption in the visible. This creates a positive feedback by reducing sea ice albedo. The process could have a more pronounced influence on global climate than the closely related carbon drawdown.

In this paper, we present results from an offline simulation which process data obtained from both atmospheric and sea ice components of the Community Earth System Model (CESM). The resulting darkening of Arctic sea ice from atmospheric deposition and biological activity will both lead to increased Arctic sea ice melt and positive climate forcing [Musilova et al. 2016]. Taking into account the possibility of increased biological activity within the Arctic Circle [Stow et al., 2004], our results show that the absorption associated with chlorophyll-a in boreal spring could be well comparable with that of black carbon under some circumstances. Specifically we find that the bottom ice layer is relatively absorptive in low latitude peripheral seas, and upper ice habitats may be dominant as black carbon plumes dilute to regions of reduced deposition. It is also noted that where soot dominates, it may appropriate photosynthetic capacity since the atmospheric source is necessarily from above.

**Model**

We developed a numerical model of a layered ice system in order to simulate light absorption contribution by the substances of interest - black carbon sedimenting from the atmosphere onto Arctic sea ice as well as chlorophyll-a generated within and below the Arctic ice pack. For consistency in setting up the comparisons, we consider both temporal and spartial distributions of the substances. This also helps to balance out any additional influence from short-term and inter-annual variation of meteorological parameters.

First, we estimate the light absorption attributable to black carbon distributed from the continents to the Arctic Ocean and other high latitude waters/ice. We rely on soot concentration distributions and deposition fields generated during the base case (IMPRV) experiment described in Qian et al. [2014] using the Community Atmospheric Model (CAM5) [Wang et al. 2013a; Qian et al. 2014]. Next we estimate chlorophyll concentrations within and lying just at the base of the ice column, mainly occurring in layers well below the depositing atmospheric carbon. For the algal calculations, we obtained our chlorophyll fields by converting bottom-ice algae nitrogen concentrations from Los Alamos Sea Ice Model (CICE) [Hunke et al., 2010], and also by constructing estimates of primary production through ocean nutrient uptake. In conclusion, we present detailed comparisons of anthropogenic and natural impacts on light absorption across the polar regime.

To maintain consistency with the IMPRV experiment, the CICE simulation was conducted on a 1 degree horizontal grid. A standard set of contemporary (year 2000) sea-ice biogeochemistry and atmospheric forcings were used to initialize the model and run for 11 years. Full meteorology and aerosol deposition were included. The first year was treated as spin up for the sea ice system so that only the final ten years were actually analyzed. We obtained monthly average ice bottom algal nitrogen concentrations (mmol/m$^2$) and estimated chlorophyll-a concentrations (mg/m$^3$) for the lowest layer using typical nitrogen to chlorophyll ratios, assuming the diatoms are dominant [Ackley et al. 1979; Geider et al. 1998; Walsh et al. 2001; Schoemann et al. 2005]. The calculations account for both upper mixed layer nutrient distributions and light limitation of algal blooms by Photosynthetically Active Radiation for algal (PAR) [Elliott et al. 2012].

Contemporary snow parameters from the CICE experiment were also used to estimate light absorption by black carbon. Our analysis for this paper is limited to boreal spring, mainly because transport and deposition of soot at polar sites are at maximum during this season [Doherty et al., 2010; Wang et al., 2013]. BC mixing as a result of sea ice/snow melt is also limited in comparison to summer season [Doherty et al., 2010]. Additionally the spring season maximizes ice biological activity. The pack matrix is still present around the periphery but snow cover and ice thicknesses are dwindling. Light and nutrients are often adequate to support strong sea ice algae blooms in this domain and production-concentration of chlorophyll is also at a maximum [Arrigo et al., 2010]. Boreal spring provides a unique opportunity to simultaneously investigate the absorption contribution of black carbon deposited onto snow alongside chlorophyll-a generated within and below the ice pack. The two phenomena can be contrasted in a spatiotemporal regime of unique climate change sensitivity.

We estimate total black carbon deposition and average column burden on the snow-ice combination for the contemporary spring period. In this study, any residual sea-ice albedo feedback that could potentially influence stability has been neglected. This is because black carbon deposited during boreal winter is minimal and in any case, there is little or no sunlight available during the polar night to drive an albedo connection [Doherty et al., 2010]. Thus, black carbon deposited in spring was allowed to mix internally with both first year and multi-year snow. A round-figure mass absorption efficiency of 10 $m^2/g$ at 550 nm was adopted, consistent with the latest black carbon review by Bond et al. (2013). Our radiation attenuation calculations are simply an application of the Beer-Lambert Law. Scattering within the column is ignored for the moment. This will later be incorporated alongside with full impact on sea ice albedo.

For chlorophyll, we considered four layers of biological activity arrayed in the vertical within sea ice. It is assumed the Arctic ice bottom layer is mainly occupied by diatoms which are not grazed significantly. The ice bottom with its enhanced porosity contains high concentration of nutrients due to its proximity to the bulk seawater column. Yet it also receives reasonable amounts of light for primary production in the spring [Ackley and Sullivan, 1994]. Habitat conditions in this layer are relatively stable, encouraging substantial productivity. The bottom layer benefits from continuous nutrient influx from the pelagic ecosystem [Horner and Alexander, 1972]. Chlorophyll concentrations adopted for this layer constitute direct estimates from our detailed CICE biogeochemistry experiment.

An Interior Layer, which forms in brine channels extending upward from the ice bottom, has sometimes been identified in columnar ice sampled from the Arctic pack [Ackley and Sullivan 1993]. Nutrients exchange into this saline environment as a function of sea ice porosity and vertical convection [Arrigo et al. 1997]. We assume that the ratio of nutrients does not change through the ice column. Vertical brine motion, which is the sole mechanism for transporting nutrient in the ice system, is only permitted when the brine volume is above a threshold value of 50 per mil [Golden et al. 1998; Vancoppenolle et al. 2010]. Nutrients are conveyed upward from the bottom layer and downward from the freeboard layer at a defined rate depending on the salinity and temperature. The Freeboard is located in the upper portion of sea ice, just at sea level. Salinity measurement in this vicinity indicates sea water diluted by ice melt. Downwelling brine and in-mixing sea water make their way into this ephemeral upper layer and may be sufficient to support algae growth rates of 0.1-1 day$^{-1}$ [Ackley and Sullivan 1994]. The process is described and portrayed in detail in Ackley and Sullivan (1994) and the resulting pigment concentration levels are verified in Arrigo et al. (1997). Special upper level habitats also form as nutrients become available in association with pressure ridging [Ackley and Sullivan, 1994] but these are of low dimensionality (fractal) and will not be discussed here.

The Infiltration Layer is situated between the snow and ice pack and can become a favorable environment for sea ice algae as a result of inflow of nutrients from the ocean through the process of snow loading and subsequent depression. We applied the Archimedes Principle and sinkage paves the way for flooding. Inflow of nutrients, ocean pigments and subsequent primary production of about 1-3 g Cm$^{-2}$ day$^{-1}$ can account for the high concentration of Chl in this zone [Ackley and Sullivan 1994]. We estimated [Chl] and optical density for upper layers and compared the light absorption capabilities with black carbon deposited on the snow. Again the potential for growth may be verified through the data in Arrigo et al. (1997). Much of the research data we rely on to study high level ice habitats must in fact be drawn from the Southern Hemisphere, because it is a fully maritime regime and precipitation rates are stronger (snow accumulation). But the processes transfer directly to the freeboard-infiltration couple in many portions of the Arctic. A key to this transfer of information: Antarctic ice strips iron from the water column so that pack biology may not be trace metal limited (Lannuzel et al. 2010).

Spring blooms were initiated when the shortwave radiation was sufficiently positive and basic nutrient restrictions were met. Nutrients fluxes to the interior and freeboard layers are treated as a function of sea surface temperature and sea ice brine volume fraction greater than 50 per mil [Arrigo et al. 2003; Vancoppenolle et al. 2010]. This is in keeping with the so-called "rule of fives" (Golden et al. 1998). Flooding at the infiltration layer was predominantly driven by

Archimedes principle while brine dynamics control the quantity of dissolved nutrient supplied to the lower communities [Reeburgh, 1984]. Here an assumption is that salt only travels vertically and there was no residual nutrient present prior to the influx. The total amount of light absorbed by the soot deposited on the ice and amounts absorbed by chlorophyll within and below the ice pack were estimated at a wavelength of 550 nm. Mass absorption efficiency of 0.03 $m^2$/mg was used to represent chlorophyll-a at this wavelength [Bannister, 1974; Suzuki et al. 1998; Yang et al. 2009]. All calculations represent attenuation as characterized by the Beer-Lambert Law. At this point, no attempt was made to incorporate ice column internal scattering.

## Results

The polar projection maps presented in this paper provide an overview of total deposited black carbon fields and chlorophyll concentrations in the vertical layers of Arctic sea ice. Although monthly averages were in fact generated for this exercise over an entire CAM year, we present output from the spring season to highlight the period when light and nutrients become available to support high chlorophyll concentrations. Files containing results for other times can be obtained from the authors on request. Regions of the Arctic Ocean lacking sea ice are blanked out in black in order to focus our attention and discussion on areas with partial or full sea ice/snow coverage. Black is likewise used to represent areas with little or none of the quantity (material) plotted. The latitude range is between 45˚ - 90˚N.

Analysis of our results shows that black carbon deposits on snow in the central Arctic at a rate of about 1 µg/$m^2$/day through the spring season (Fig. 1, upper left). This value is small by comparison with amounts accumulating at lower latitudes near the ice edge. Deposition around Sea of Okhotsk in Russia and areas surrounding North Sea and Baltic in Europe experience orders of magnitude greater sedimentation, due to their proximity to strong point sources [Bond et al., 2010]. As a North American reference, mixing ratios of black carbon in snow on the Bering Sea and Hudson Bay are in the range 15-20 nanogram per gram (Fig. 1, lower left). As expected, regions with substantial soot deposition in the Arctic show reduced transmittance of solar radiation passing through the snow/ice column (Fig. 1, lower right). A significant absorption of radiation in this region could lead to enhanced snow melt. By contrast the Canadian Archipelago and central Arctic are high-transmittance low-absorption areas so that visible radiation passes through the snow with little or no absorption.

Plots of chlorophyll concentrations and optical depth for the four different vertical (biological) layers within Arctic sea ice are presented in Figure 2(a – d) starting sequentially from the uppermost or infiltration layer then moving downward toward the ice bottom. Both the infiltration and freeboard habitats reside nearer the light source so that blooms can begin as soon as nutrient restrictions are surmounted. At the infiltration layer, ocean nutrients transported onto ice-snow interface produce more [chl] near the coastal ice-edge than land ice-edge. This effect can be attributed to our inclusion and treatment of Archimedes Law relative to snow accumulation. For example, concentrations computed for infiltration layers in the maritime regime at the Sea of Okhotsk and Bering are one order of magnitude greater than for the Kara Sea or the Siberian Shelf (Fig. 2a, left). In some ecogeographic zones across the Arctic, the freeboard layer tends to be partially absent or without any significant [chl] in the boreal spring (Fig. 2b, left).This is due to a combination of low air temperatures with the requirements of the rule of 5's. Visible radiation absorption by chlorophyll in the two uppermost layers of biological activity is quite comparable (Fig. 2a right and 2b right).

Biomass of the interior ice algal community is limited at this time because growth is still restricted in the column by low porosity. Higher concentrations are only apparent at the ocean-ice boundary where porosity is elevated. The entire central Arctic lacks conditions suitable for significant chlorophyll production below the freeboard. In our model this is represented through low internal temperatures and their impact on brine thermodynamics. If an interior habitat is present at all, concentrations for areas north of 75°N and in the Canadian Archipelago were less than 0.1 µg/m$^3$ (Fig. 2c, left). The highest chlorophyll concentration estimated for this zone surround the Sea of Okhotsk, where concentration reaches about 0.2 mg/m$^3$. The interior has the lowest chlorophyll concentration of all four layers considered, so that solar absorption due to the pigment is quite limited. Note that the thickness assigned to this habitat for purposes of the optical calculations conducted here is 50% of the entire ice thickness.

As the sun rises higher in the polar sky in the spring, the sea ice bottom layer with its strong nutrient supply begins to receive enough photosynthetically active radiation (PAR) to initiate photosynthesis. Blooms can last for many weeks, until surface melt leads to flushing (Lavoie et al. 2005; Jin et al. 2006). Chlorophyll production may thus rise steadily to a late maximum. Concentrations at the ice bottom and upward for several centimeters accumulate to high values over the entire region covered by sea ice (Fig. 2d, left). In our simulations, levels peaked at about 300 – 1000 mg/m$^3$ in the Bering and Sea of Okhotsk. The bloom at the bottom layer in the central Arctic could produce chlorophyll concentrations in the tens of mg/m$^3$. Optical depth due to the presence of chlorophyll in this layer could be quite large taken on a relative basis (Fig. 2d, right). Significant absorption of downwelling radiation should thus occur – but it must be recalled that bottom habitats sit several tens of centimeters below the ocean and ice surface as a general rule. Pure sea ice itself has an attenuation coefficient of about 1 per meter (Lavoie et al. 2005). An e-fold or so must be expected prior to photons reaching this intense zone of biological activity. You may notice in 2d that there are several light blue areas on the Atlantic side perimeter. These are due to and reflect fractional grid cell ice coverage as tracked in the CICE physics code.

Finally, we compared the optical depth computed as a result of black carbon deposited on Arctic snow/ice with values generated as a result of algal chlorophyll, at the three column positions with highest concentration. Comparisons involving the freeboard and infiltration layers are presented together here, since attenuations are similar even though geographic distributions are distinct. The interior is set aside at this point. While the existence of ecosystems deep inside the center of the pack is of considerable biological interest, we conclude that absorption is modest. It is clear from our effort that within the entire Arctic Circle, black carbon deposited on snow in boreal spring possesses higher optical depth and will thus absorb a greater proportion of incident, downwelling solar radiation. We focus now on the column integral of black carbon, along with the ice top and bottom.

The system of ecogeographic zones adopted in this work to organize our analysis is illustrated in the upper left panel of Figure 3. In the remainder of the figure, selected sets of optical depths are rationed to one another then log-transformed. In most regions of the Arctic, high chlorophyll concentration at the bottom layer of Arctic sea ice will have an optical depth which is several orders of magnitude higher than estimates for upper layers in the spring. However at the ice-ocean perimeter along Greenland and in some parts of the Bering Sea, the seasonal melt has already begun and suppresses activity low in the column through purging (Lavoie et al. 2005; Jin et al. 2006 Fig. 3, top right). We do not purge the infiltration layer in the current computations, since portions of it may reside above drainage channels. In comparison with black carbon, the chlorophyll generated by ice algal at the ice bottom is sometimes orders of magnitude absorptive (Fig. 3, lower left). Low level purging again leads to low ratios around the ice edge. However, in the Barents Sea, soot deposition is still dominant. In a further comparison, light absorption strength by the carbon exceeds that of infiltration chlorophyll over the entire Arctic except at small areas in the Baffin Bay (Fig. 3, lower right). The effect tends to decay moving toward the pole since pollution sources become more remote.

## Model Validation

Chlorophyll concentrations presented in this paper were validated against an ice algal chlorophyll data base developed for optimization of the general sea ice biogeochemistry code [Elliott et al. 2016]. Generally, the visual validation shows a gradual increase in phytoplankton bloom towards the spring season. As the bloom peak following the sun from winter to spring, the bottom layer tends to maximize at 300 to 3000 mg/m$^3$. Upper level habitat data, while rare, ranged from 0 to 1 in the same units moving outward from the pole [Thomas et al. 1995; Gosselin et al. 1997; Smith et al. 1997; Gradinger 1999; Rysgaard et al. 2001; Uzuka et al. 2003; Lavoie et al. 2005]. Our black carbon deposition calculations were similarly validated as described in Qian et al. (2013) and Wang et al. (2013).

## Discussion

Arctic sea ice extent is actually disappearing at a faster rate than predicted by model simulations. Light absorption by black carbon and other associated processes currently presented in climate model seems to be insufficient to account for the notable disappearance of Arctic sea ice observed in recent times. Thus, it is imperative to investigate the degree of contribution of some of the biogeochemically derived light absorbing materials, such as chlorophyll, in diminishing Arctic sea ice extent. As the sea ice extent continues to thin, there is high tendency for increased light penetration which concomitantly drives primary production within the ice [Stow et al., 2004; Nicolaus et al., 2012], ultimately leading to an increased chlorophyll concentration and further light absorption.

Analyses of results in the current work show that spring time BC total deposition diminishes to about 1 µg/m$^2$/day as the material migrates from the point sources toward the central Arctic. Similar trend, as expected, is obvious in the optical depth. However, chlorophyll concentrations beneath and within the ice pack do not follow any designed pattern. Concentrations at the sea ice bottom layer in some areas of the central Arctic are almost 300 mg/m$^3$. The biopigment tends to have considerable to more light absorption as we move away from the poles towards lower latitudes because of the presence of an expanding melt – pond coverage resulting from first year ice. Other layers of Arctic sea ice also show significant increase in chlorophyll concentrations at ice – land edges as the ice thins from the edges.

In the boreal spring, light absorption by chlorophyll present in certain layers of Arctic sea ice may thus be significant when compared to the contribution of anthropogenic BC. Some biological habitats including the bottom and infiltration layers are quite absorptive, approaching or surpassing the value for black carbon. During the contemporary period, it is suspected that thinning sea ice will lift radiation restrictions and allow biological activity to increase. Meanwhile black carbon point sources are being restricted under new environmental regulations in most developed countries at the Northern Hemisphere [Bond et al., 2010].

For other seasons, when the light and nutrient regimes are more restrictive, black carbon deposited on snow may be broadly responsible for light absorption. There is also the issue of penetration through the multiple layers of snow and ice. In this current study, we have limited the discussion here to the simple quantity optical depth to focus on relative attenuation moving downward into the sea ice system. A next step will be to move to one dimensional radiation transfer calculations, likely of the simple two stream variety. It will then be possible to compute local energy deposition and more from relative to absolute analysis. Since the attenuation coefficient of standard sea ice is about 1 per meter, we speculate that there will be significant penetration and then biological absorption for thicknesses of this order or less. As Arctic ice coverage becomes more tenuous in coming decades, it seems possible that chlorophyll will be a major driving force in the light absorption and the energy budget of the polar cap. We anticipate the introduction of competing soot and chlorophyll absorption calculations into polar systems models in the near future.

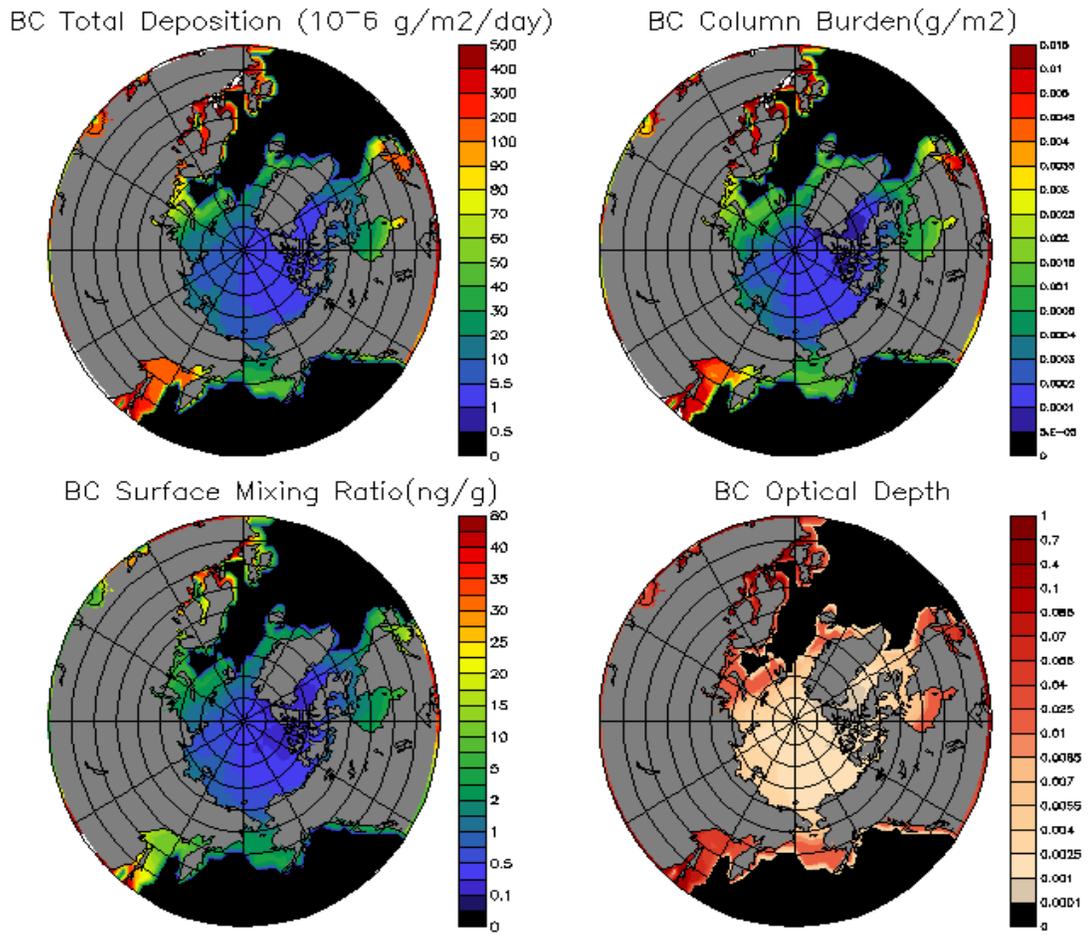

**Figure 1:** Spatial distribution of Black Carbon (BC) deposited on snow-ice (during spring season - MAM) absorbs significantly in parts of the Arctic

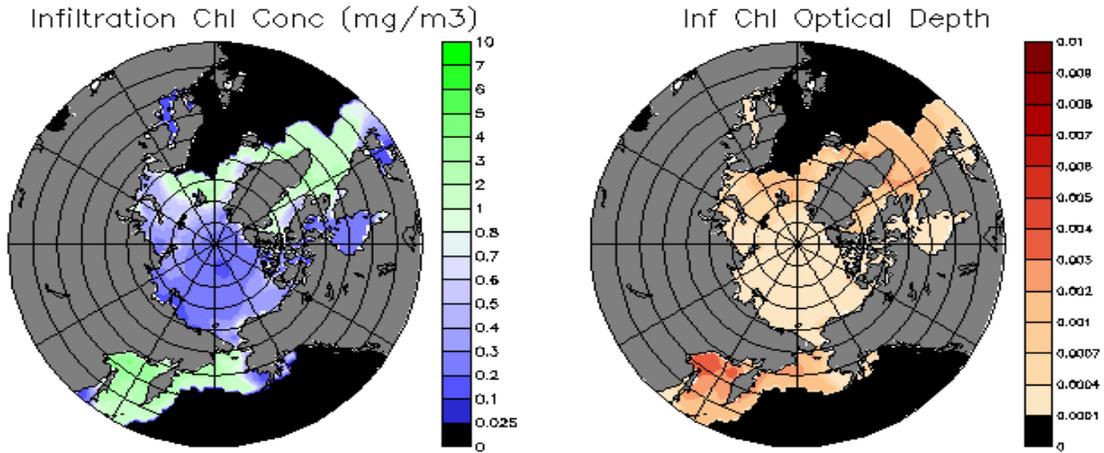

**Figure 2a:** Spring chlorophyll concentration and optical depth, as light passes through the Infiltration (Inf) layer of Arctic Sea Ice.

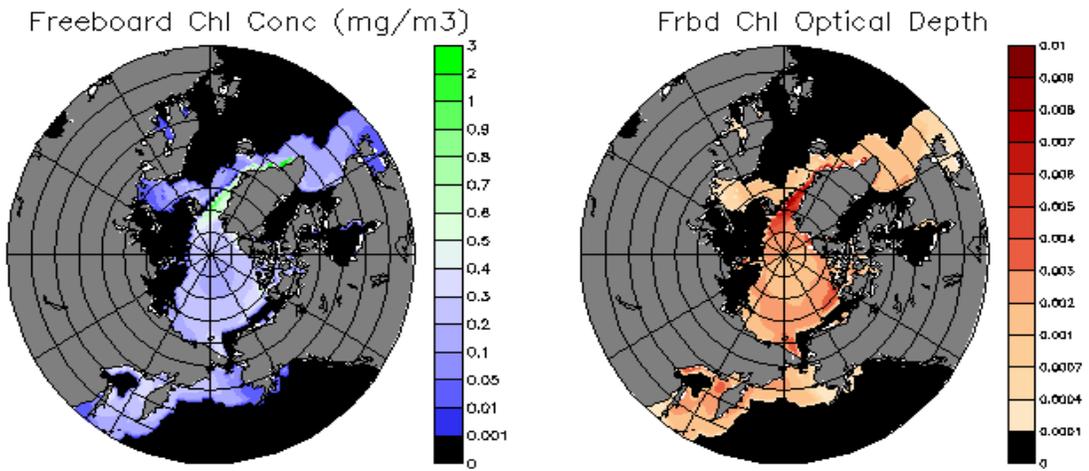

**Figure 2b:** Spring chlorophyll concentration and optical depth, as light passes through the freeboard (Frbd) layer of Arctic Sea Ice.

**Figure 2** continued

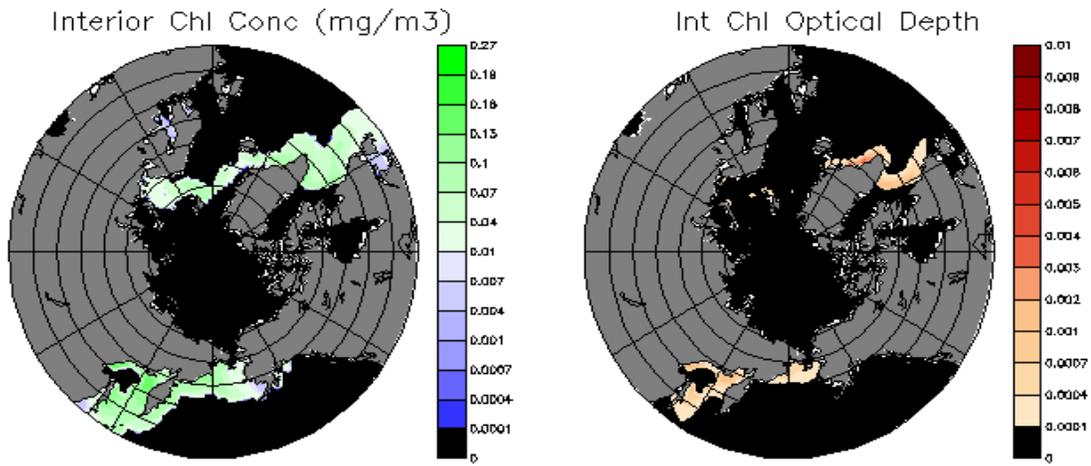

**Figure 2c:** Spring chlorophyll concentration and optical depth, as light passes through the Interior layer (Int) of Arctic Sea Ice

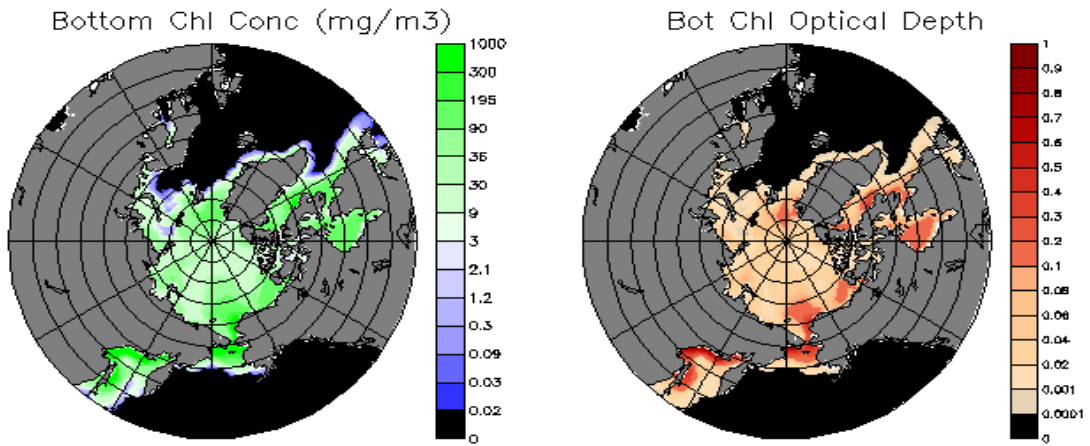

**Figure 2d:** Spring chlorophyll concentration and optical depth, as light passes through the bottom (Bot) layer of Arctic Sea Ice

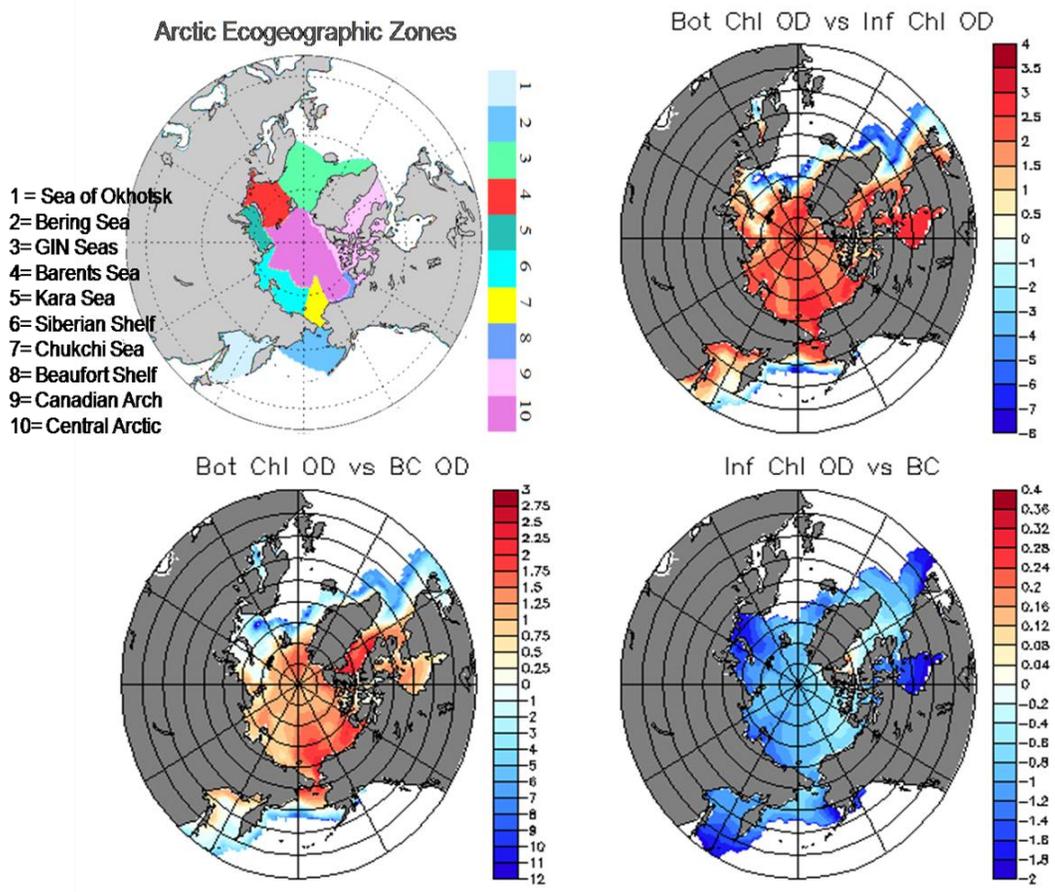

**Figure 3:** Eco-zones then examples for Arctic Sea ice optical depths, ratioed in permutation to the corresponding values for black carbon and chlorophyll, given as base10 logarithms. Bot = Bottom Layer, Inf = Infiltration Layer, BC = Black Carbon and Chl = Chlorophyll.

**Acknowledgements**

Participants at Los Alamos National Laboratory and New Mexico Tech thank the U.S. Departmentof Energy SciDAC program (Scientific Discovery for Advanced Computing), and specifically its ACES4BGC project (Applying Computationally Efficient Schemes for Biogeochemical Cycles). OO and OW were supported by Los Alamos Institute of Geophysics, Planetary Physics and Signature (IGPPS), now Center for Space and Earth Science (CSES). We also appreciate Jeff Altig and Forrest Hoffman for the use of the modeling lab in New Mexico Tech and Oak Ridge National Laboratory supercomputer respectively.